\documentclass[sn-nature]{sn-jnl}

\usepackage{graphicx}%
\usepackage{multirow}%
\usepackage{amsmath,amssymb,amsfonts}%
\usepackage{amsthm}%
\usepackage{mathrsfs}%
\usepackage[title]{appendix}%
\usepackage{xcolor}%
\usepackage{textcomp}%
\usepackage{manyfoot}%
\usepackage{booktabs}%
\usepackage{algorithm}%
\usepackage{algorithmicx}%
\usepackage{algpseudocode}%
\usepackage{listings}%
\usepackage{graphicx}
\usepackage{amsmath}
\usepackage{amssymb}
\usepackage{booktabs}

\usepackage{wrapfig}
\usepackage{multirow}
\usepackage{bbding}
	
\usepackage{color, colortbl}

\definecolor{Gray}{gray}{0.9}
\newcolumntype{a}{>{\columncolor{Gray}}c}

\begin{document}

\title{Accurate Patient Alignment without Unnecessary Imaging Dose via Synthesizing Patient-specific 3D CT Images from 2D kV Images }


\author[1]{Yuzhen Ding, PhD}
\author[1]{Jason M. Holmes, PhD}
\author[1]{Hongying Feng, PhD}
\author[2]{Baoxin Li, PhD}
\author[1]{Lisa A. McGee, MD}
\author[1]{Jean-Claude M. Rwigema, MD}
\author[1]{Sujay A. Vora, MD}
\author[3]{Daniel J. Ma, MD}
\author[3]{Robert L. Foote, MD}
\author[1]{Samir H. Patel, MD}
\author[1]{Wei Liu, PhD\thanks{corresponding author: Liu.Wei@mayo.edu}}

\affil[1]{Department of Radiation Oncology, Mayo Clinic, Phoenix, Arizona, 85054}
\affil[2]{School of Computing and Augmented Intelligence, Arizona State University, Phoenix, Arizona, 85287}
\affil[3]{Department of Radiation Oncology, Mayo Clinic, Rochester, Minnesota, 55905}

\date{}


\abstract{In radiotherapy, 2D orthogonally projected kV images are used for patient alignment when 3D-on-board imaging(OBI) unavailable. But tumor visibility is constrained due to the projection of patient’s anatomy onto a 2D plane, potentially leading to substantial setup errors. In treatment room with 3D-OBI such as cone beam CT(CBCT), the field of view(FOV) of CBCT is limited with unnecessarily high imaging dose, thus unfavorable for pediatric patients. A solution to this dilemma is to reconstruct 3D CT from kV images obtained at the treatment position. Here, we propose a dual-models framework built with hierarchical ViT blocks. Unlike a proof-of-concept approach, our framework considers kV images as the solo input and can synthesize accurate, full-size 3D CT in real time(within milliseconds). We demonstrate the feasibility of the proposed approach on 10 patients with head and neck(H\&N) cancer using image quality(MAE:$<$45HU), dosimetric accuracy(Gamma passing rate(2\%/2mm/10\%):$>$97\%) and patient position uncertainty(shift error:$<$0.4mm). The proposed framework can generate accurate 3D CT faithfully mirroring real-time patient position, thus significantly improving patient setup accuracy, keeping imaging dose minimum, and maintaining treatment veracity.}

\maketitle

\section{Introduction}\label{sec1}
Radiotherapy (RT) is a standard and a favored treatment modality for head and neck (H\&N) cancer patients. Among the cutting-edge approaches in RT, intensity-modulated proton therapy (IMPT) distinguishes itself by precisely delivering maximum cell-killing energy to tumors while minimizing exposure to surrounding organs at risk (OARs)\cite{schild2014proton,liu2012influence,liu2015impact,stuschke2012potentials}. Despite its precision, IMPT remains highly susceptible to factors such as patient setup, proton beam range uncertainties, respiratory motion, and inter-fractional anatomical changes\cite{lomax2008intensity,deng2021critical,liu2018small,matney2013effects,matney2016perturbation,quan2013preliminary,shan2018robust}. These uncertainties may potentially result in undertreatment of tumors or excessive exposure of surrounding OARs, leading to local recurrence and unexpected treatment-related adverse events (AEs)\cite{ang2010human, yang2021exploratory, yang2022exploratory, yang2022empirical}.

Image-guided patient alignment is an essential step for patient setup in RT, especially for the more uncertainty-vulnerable IMPT. During treatment, before delivering the prescribed dose, the therapists need to carefully adjust the patient's position and posture to align with the planning CT, on which the treatment plan was designed. Hence, accurate dose delivery depends on accurate patient setup, for which minimizing patient alignment error is vital. In RT, commonly employed on-board imaging (OBI) techniques are CT-on-rails (CToR), cone beam CT (CBCT), and orthogonal kV images, etc. CToR has the diagnostically equivalent imaging quality (same as planning CT) but requires effective transfer of the patient from the CT scanner to the position of treatment. During the transfer, any uncertainty caused by either patients' movement or position/posture discrepancy can negate the effectiveness of image guidance. CBCT is used at the treatment position but has artifacts that could produce greater dosimetric errors in the correspondingly calculated dose distribution in the patient. Meanwhile, CToR and CBCT are relatively expensive, which impedes their widespread adoption, especially in low-income, rural areas. In addition, the rather high imaging doses from both CToR and CBCT act as a barrier to frequent use (especially for pediatric patients), such as the promising daily online adaptive radiation therapy (ART)\cite{yan1997adaptive,yan1998use, feng2022gpu}. A recent study\cite{kaushik2024generation} proposed to synthesize virtual 3D CT from CBCT utilizing the commercially available software, partially addressing the drawbacks of CToR (additional alignment uncertainty due to indispensable position transfer and posture change) and CBCT (low image quality). However, the imaging dose is still not reduced and the process was complicated such that any errors accumulated during the procedure may ultimately lead to significant setup errors. Orthogonal kV image performs in a real-time manner with much less expense and a much lower imaging dose compared to CToR and CBCT. Nevertheless, image quality-wise, kV image only shows clear bony anatomies in a 2D x-ray projection(e.g. the middle image in Fig. \ref{fig:orthogonal_xray_system}), making it barely usable in online ART workflow except for patient alignment. Even for patient alignment, these 2D images often lack sufficient soft tissue details, leading to large patient setup uncertainties. The ad hoc handling of large patient setup uncertainties by using large target margins can lead to unnecessarily high dose to nearby OARs and thus unnecessary AEs. Overall, there is currently no imaging guide technique for radiation therapy that can simultaneously be real-time in efficiency, diagnostically equivalent in quality, and cheap and low dose for frequent and widespread use. 

\begin{figure}[!th]
  \centering  
  \includegraphics[width=1\linewidth]{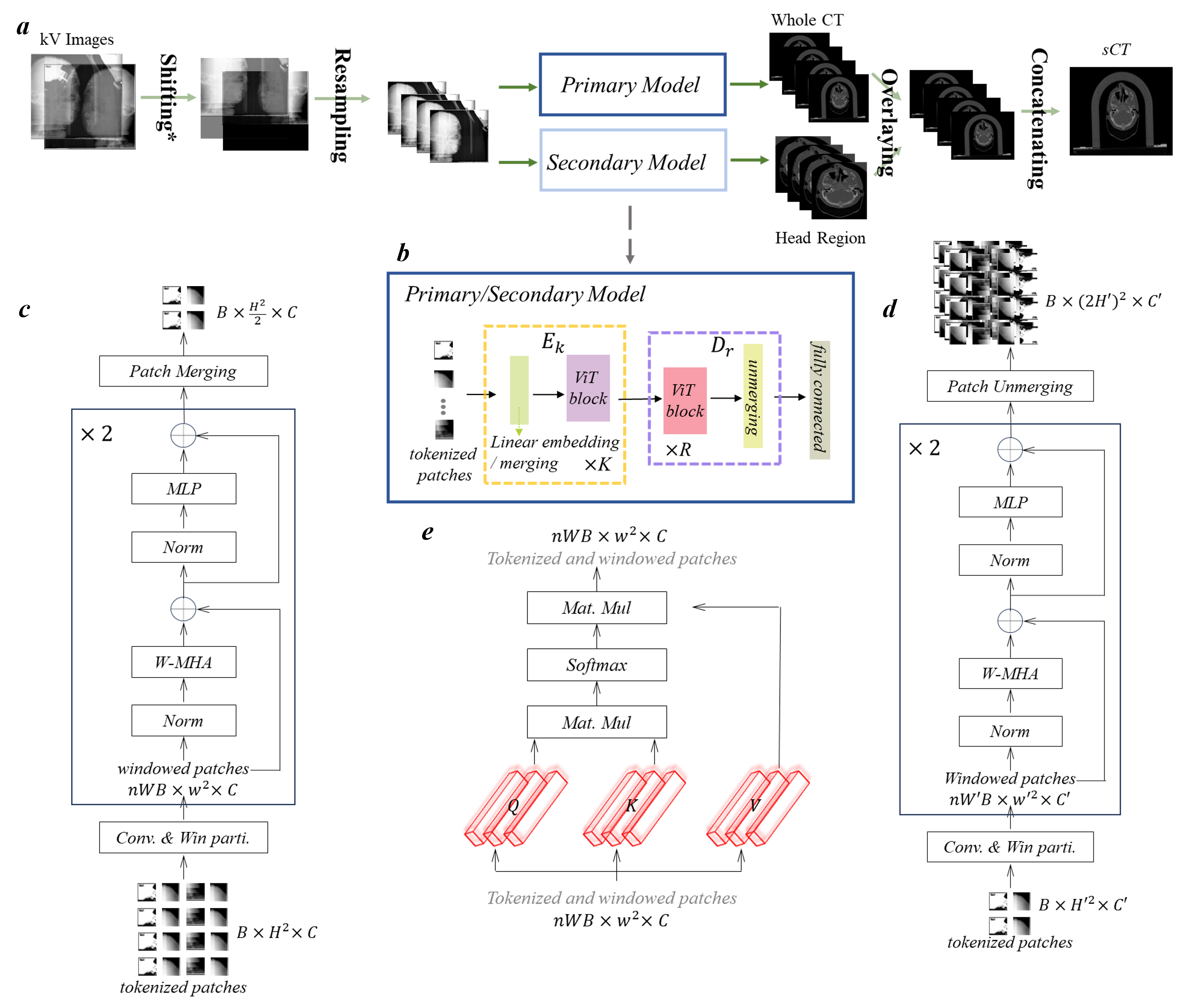}
   \caption{Overall architecture of kV2CTConverter.\textbf{\textit{a.}} Workflow of the proposed method. The raw kV images were augmented by GRSS to get adequate samples for model training. Then the processed images simultaneously went through dual models (i.e., primary model and secondary model) to generate the whole CT and the fractional CT that covered only the head region, respectively. Lastly, the full-size synthesized CT was achieved by overlaying and concatenating the outputs from two models according to their spatial relationship. \textbf{\textit{b.}} The model structure of both primary and secondary model.\textbf{\textit{c.}} The details of the hierarchical ViT blocks in the encoder $E_k$. \textbf{\textit{d.}} The details of the hierarchical ViT blocks in the decoder $D_r$. \textbf{\textit{e.}} The detailed illustration of the window-based Multi-Head Attention (W-MHA), the tokenized patches were first spat to $nW$ non-overlapped windows of a size of $w \times w$ and the attention was only calculated on the windows instead of the whole inputs.}
   \label{fig:model structure}
\end{figure}

Recently, artificial intelligence (AI) has undergone rapid development and its application in radiation oncology has been growing quickly\cite{LIU2023100045, dingregistration2023, BALAGOPAL2021102195,jiang20223d}. The feasibility of using AI models to inversely reconstruct the 3D CT from 2D clean and noise-free digital reconstructed radiography (DRR) images (e.g. the right image in Fig. \ref{fig:orthogonal_xray_system}) has been explored and validated\cite{shen2019patient,ying2019x2ct,bayat2020inferring,lei2020deep, maken20232d,gao20233dsrnet,zhang2023xtransct}. Though conceptually inspiring, such a model itself is chronologically inapplicable in clinical practice since the DRR image is generated based on the planning CT or CToR images previously obtained. Only independent 2D images without prerequisite 3D images, such as kV images, are the practically meaningful inputs to such 2D-to-3D deep learning-based models. In addition, most existing deep learning-based 3D CT reconstruction approaches are tested and validated on small-size images only (typically $128 \times128 \times100$), which are far from clinically satisfactory. Therefore, we propose a novel deep learning-based framework kV2CTConverter (Fig. \ref{fig:model structure}\textbf{\textit{a}}) composed of dual models built with hierarchical vision transformer (ViT) blocks\cite{dosovitskiy2020image,liu2021swin}. The proposed framework will take kV images as the exclusive input to synthesize the corresponding full-size 3D CT in a real-time manner, which can be used for reflecting real-time patient's position in 3D, thus achieving high-quality but almost "zero-dose" image-guided patient alignment. To our best knowledge, kV2CTConverter is the first AI-based approach that utilizes only clinically available images (daily kV images) and reconstructs real-time, ready-to-use, full-size 3D CT, paving the way towards AI-based online ART. We believe this work will have broad impact upon image-guided interventional procedures such as radiation therapy and needle biopsy, especially for pediatric patients who are sensitive to imaging dose. In addition, it may help simplify the hardware of tomographic imaging systems.  Also, the developed real-time CT reconstruction with limited projections can benefit radiation therapy in low-income, rural areas, where many radiotherapy machines lack 3D OBI capability.

 
 At the core of bridging the theoretical algorithm and the practical application, a hierarchical vision transformer is adopted and adapted to the medical images (i.e., kV images and 3D CT) with a dual-model setting and a data augmentation strategy termed as geometry property reserved shifting and sampling (GRSS) is also proposed. GRSS is a novel yet easy-to-implement data augmentation method that takes the geometrical relation between the treatment couch and kV imaging source and detector into consideration. This new strategy enables kV2CTConverter to take full advantage of the noisy but ultra-sparse 2D kV images to fulfill accurate 3D CT reconstruction while avoiding model overfitting. We validated the effectiveness of the kV2CTConverter using 10 independent patients with H\&N cancer from three perspectives: 1) image quality evaluation of the synthesized CTs, 2) dosimetric evaluation of the dose distributions calculated using the synthesized CTs, and 3) robustness of the framework to random shifts that mimic patient alignment uncertainties during treatment. In both image quality and dosimetric evaluations, kV2CTConverter exhibited high accuracy. In random shift test, the kV2CTConverter achieved a minimum shift error of 0.4mm, which meets the clinical criterion.

\section{Results}\label{sec2}
\subsection{Image quality of the reconstructed 3D CT} 
We evaluated the synthetic CT (sCT) with mean absolute error (MAE) and absolute per-voxel CT numbers (i.e., image intensity in terms of Housfield Unit (HU)) difference volume histogram (CDVH), compared to the CToR that was taken on the same day as the input kV images (i.e. ground-truth CT (gCT)). 3D Gamma analysis of the CT numbers comparing the sCT and gCT was done where the calculation was done twice, with sCT and gCT as the reference CT, respectively. The second column in Table \ref{MAE results} listed the results from 10 patients in terms of MAE (in HU).  On average, the proposed framework achieved an MAE of $44.58 \pm 20.21$, indicating the sCT and gCT yielded a high agreement at the voxel level. Compared to the numerical results reported in \cite{lei2020deep,shen2019patient}, although theirs were calculated on small-size 3D lung CT (supposed to be easier as the structures were much larger and had more distinct boundaries), ours still outperformed them by 53.56\% and 62.01\% on average, respectively.

\begin{figure}[h]
  \centering  
  \includegraphics[width=1\linewidth]{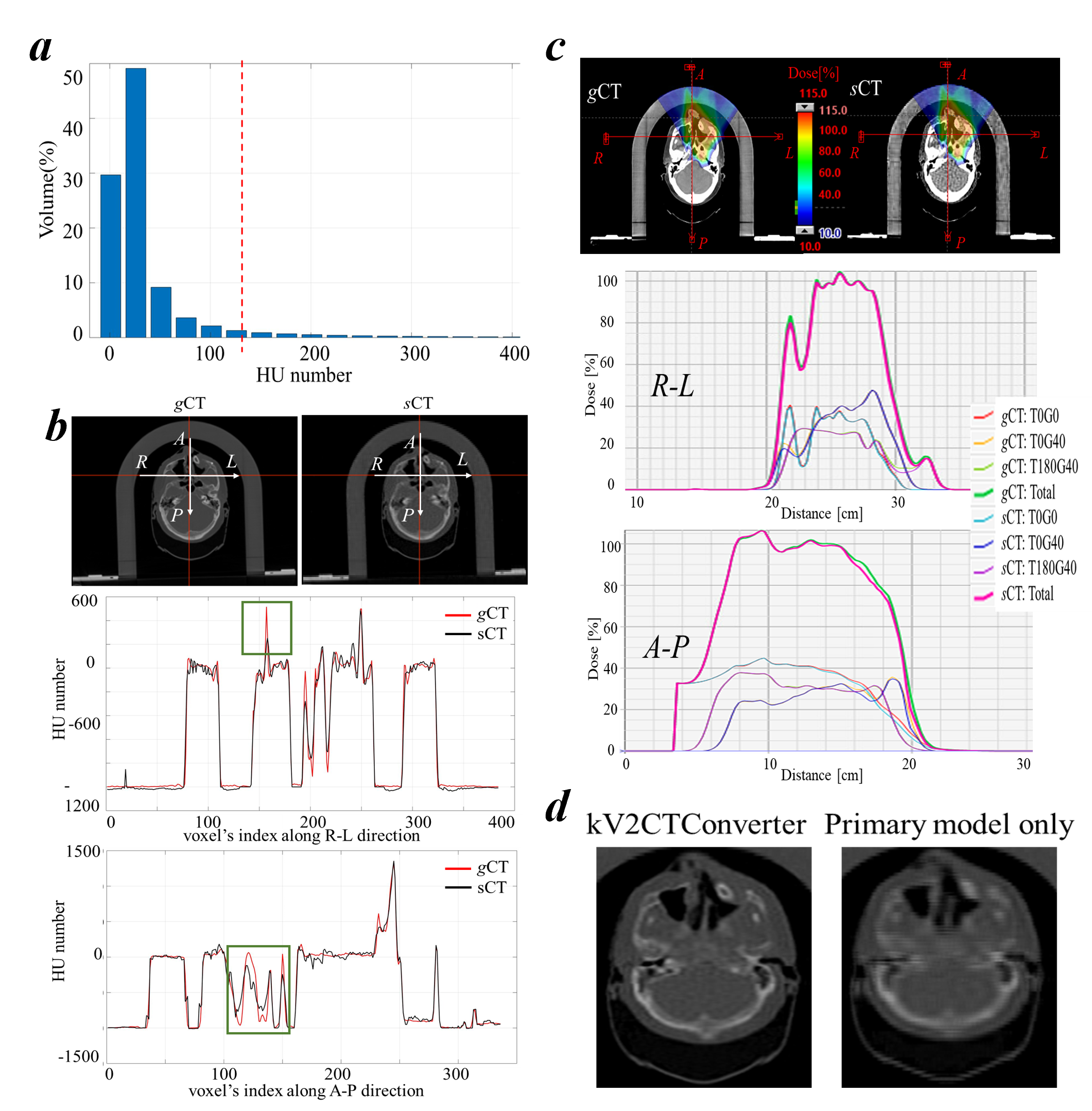}
   \caption{Experimental results from a typical patient.\textbf{\textit{a.}} CDVH of the sCT from a typical patient \textbf{\textit{b.}}HU number profile comparison between sCT and gCT in both R-L and A-P directions.The green box showed the area where the discrepancy was large.  \textbf{\textit{c.}}Dose profile comparison between the doses calculated on the sCT and on the gCT in both both R-L and A-P directions.\textbf{\textit{d.}} Enlarged head region comparison between the kV2CTConverter and primary model only.}
   \label{fig:results}
\end{figure}

Fig.\ref{fig:results}\textbf{\textit{a}} depicted the CDVH of one typical patient. It is clear that the majority of the voxels had a per-voxel CT number absolute difference of less 100 HU. Numerically, only 5\% of the voxels had an absolute difference larger than 128HU. We also randomly selected one slice from the 3D CT image of one typical patient and conducted the HU number profile comparison to visualize the difference in both right-left (R-L) and anterior-posterior (A-P) directions between the gCT and sCT in Fig.\ref{fig:results}\textbf{\textit{b}}, where the red curve represented the gCT while the black one represented the sCT. We observed that the two curves were highly overlapped, showing the sCT matched well with gCT. The 3D Gamma passing rates of comparing sCT to gCT for all ten patients were calculated and the results were reported in Table \ref{ctgamma} and Table \ref{ctgammafull}. Regardless of reference CTs selected for calculation, our proposed approach achieved 3D Gamma passing rate of $98.05\% \pm 1.82\%$, $97.79\% \pm 1.96\%$ and $97.79\% \pm 1.96\%$ on average with criteria of 3\%3mm, 3\%2mm and 2\%2mm, respectively. 

\renewcommand{\arraystretch}{2}
\begin{table}[h]
  \centering
    \caption{MAE (in HU) results from all 10 test patients. The lower the MAE value is, the better the image quality is. The second column showed the results from kV2CTConverter while the third column was the results from only the primary model. The last column was the difference between the kV2CTConverter and the primary model only.}
    \label{MAE results}    
  \begin{tabular}{ cacc }
  \toprule
    {\bf Patient ID} &{\bf kV2CTConverter} &{\bf Primary model only} & {\bf Difference} \\
    \hline
    \hline
    {\bf 1}&{30.19} &{34.47}&{-4.28}\\
    {\bf 2}&{28.93} &{34.00}&{-5.07} \\
    {\bf 3}&{75.14}& {88.53}& {-13.39}\\
      {\bf 4} &{40.44}&{47.85}& {-7.41}\\
      {\bf 5}&{33.76}& {38.44}&{-4.68}\\
      {\bf 6}&{39.83}&{41.81}&{-1.98}\\
      {\bf 7}&{34.34}& {38.56}&{-4.22}\\
       {\bf 8}&{46.78}&{50.34}&{-3.56}\\
       {\bf 9}&{86.98}&{99.05}&{-12.07}\\     
       {\bf 10}&{29.45}&{35.47}&{-6.02}\\
       {\bf average}&{44.58 $\pm$ 20.21}&{50.85 $ \pm$23.40}&{-6.27 $ \pm$3.71}\\
      \bottomrule
  \end{tabular} 
\end{table} 


\begin{table}
  \centering
    \caption{3D Gamma Passing rates (in \%) comparing the CT numbers (HU) of sCTs to gCTs. Three criteria were considered with sCT and gCT as the reference, respectively. For each model, sCT-referenced and gCT-referenced average results of the 10 patients were reported in the first row while the forth and last rows showed the average values of the 10 patients regardless the reference selected for calculation. The detailed results of the 10 patients were reported in Table \ref{ctgammafull} in Appendix \ref{secA1}. }
    \begin{tabular}{ccaacc}
    \toprule
    \multicolumn{2}{c}{\bf{Model}} &
       \multicolumn{2}{c}{ \bf{kV2CT}}& \multicolumn{2}{c}{ \bf{Primary}}\\\hline
        \multirow{2}{*}{3\%3mm}&sCT& 97.39 $\pm$ 2.09  &&97.29 $\pm$ 1.69&\multirow{2}{*}{97.23 $\pm$ 1.80}\\
        &gCT& 98.70 $\pm$ 1.27 &\multirow{-2}{*}{98.05 $\pm$ 1.82}&97.17 $\pm$ 2.00&\\
        \multirow{2}{*}{3\%2mm}&sCT& 97.05 $\pm$ 2.24 &&96.74 $\pm$ 1.91&\multirow{2}{*}{96.89 $\pm$ 1.91}\\
        &gCT& 98.53 $\pm$ 1.36 &\multirow{-2}{*}{97.79 $\pm$ 1.96}&97.04 $\pm$ 2.01&\\        
        \multirow{2}{*}{2\%2mm}&sCT& 97.04 $\pm$ 2.24 &&96.69 $\pm$ 1.97&\multirow{2}{*}{96.86 $\pm$ 1.95}\\
        &gCT& 98.53 $\pm$ 1.36 &\multirow{-2}{*}{97.79 $\pm$ 1.96}&97.03 $\pm$ 2.02&\\ 
    \bottomrule 
  \end{tabular}
  \label{ctgamma}
\end{table}
  
\subsection{Dosimetric evaluation of the reconstructed 3D CT}
We conducted forward dose calculation on both the sCTs and gCTs using the same plan and compared the dose distributions using 3D Gamma analysis. The corresponding dose volume histogram (DVH) indices of targets and OARs were compared as well. Fig. \ref{fig:results}\textbf{\textit{c}} depicted a typical dose profile comparison between the doses calculated on the sCT and the dose calculated on the gCT in both R-L and A-P directions. We found that the dose calculated on the sCT was very close to that calculated on the gCT for every treatment field and all fields accumulated.  Table \ref{dosegamma} showed the 3D Gamma passing rate results where the calculation was done twice, with the dose calculated on the sCTs and the dose calculated on the gCTs as the reference dose, respectively. Our framework achieved high passing rates of 99.66\% $\pm$ 0.39\%, 98.93\% $\pm$ 1.15\%, and 97.85\% $\pm$ 1.76\% on average with criteria of 3\%3mm, 3\%2mm, and 2\%2mm, respectively. Table \ref{dvhdiff} showcased the difference in DVH indices of CTV and 4 OARs between the doses calculated on the sCTs and the gCTs using the same plans, respectively. It is worth noting that only the DVH index differences were reported as the plans and dose volume constraints varied from patient to patient. Hence, DVH index difference was considered to be a better sCT-vs-gCT similarity indicator than the DVH index absolute value itself. We noticed that the proposed framework achieved a negligible difference for D95\% and D2\% of CTV, indicating the equivalence of CTV coverage and hot spot control in both dose distributions. For the selected OARs, the DVH index differences were also small. All these results showcased that the doses calculated on the sCTs and gCTs using the same plan exhibited minimal disparity, indicating that sCT could serve as a viable substitute for verification CT for the purposes of plan evaluation or adaptive re-planning.


\begin{table}
  \centering
    \caption{3D Gamma Passing rates (in \%) of the doses calculated on the sCTs and the gCTs using the same plan. Three criteria were considered with the dose calculated on sCT and gCT as the reference, respectively. For each model, sCT-referenced and gCT-referenced average results of the 10 patients were reported in the first row, while the forth and last rows showed the average value of the 10 patients regardless the reference selected in the calculation. The detailed results of the 10 patients were reported in Table \ref{dosegammafull} in Appendix \ref{secA1}}
\begin{tabular}{ccaacc}
    \toprule
    \multicolumn{2}{c}{\bf{Model}} &
       \multicolumn{2}{c}{ \bf{kV2CT}}& \multicolumn{2}{c}{ \bf{Primary}}\\\hline
        \multirow{2}{*}{3\%3mm}&sCT&99.67 $\pm$ 0.34 & & 94.11 $\pm$ 2.57 & \multirow{2}{*}{93.57 $\pm$ 2.93}\\
        &gCT &99.65  $\pm$ 0.45&\multirow{-2}{*}{99.66 $\pm$ 0.39} &93.04 $\pm$ 3.30&\\
        \multirow{2}{*}{3\%2mm}&sCT&98.95 $\pm$ 1.08 && 90.32 $\pm$ 3.84& \multirow{2}{*}{90.01 $\pm$ 4.033}\\
        &gCT &98.91 $\pm$ 1.27&\multirow{-2}{*}{98.93 $\pm$ 1.15} &89.69 $\pm$ 4.39&\\
        \multirow{2}{*}{2\%2mm}&sCT& 97.87 $\pm$ 1.71 & & 87.59 $\pm$ 4.83  & \multirow{2}{*}{87.12 $\pm$ 5.00}\\
        &gCT &97.82 $\pm$ 1.89&\multirow{-2}{*}{97.85 $\pm$ 1.76} &86.65 $\pm$ 5.39&\\
    \bottomrule    
   
  \end{tabular}
  \label{dosegamma}
\end{table}

\subsection{Shift robustness evaluation: mimicking patient setup uncertainty} 
As the primary application of the proposed framework is to generate accurate 3D CT for patient alignment during RT treatment, we conducted a comprehensive analysis of the robustness of the proposed framework to generate sCTs in the face of patient setup uncertainties. We performed random shifts to simulate patient setup uncertainty. Given a manually shifted kV images within $\pm$ 4.5mm as input, on one hand the model predicted the shifted sCT(ssCT). On the other hand, the shifted gCT (sgCT) could be calculated based on the geometrical relation between the treatment couch and kV imaging system (see Fig. \ref{fig:orthogonal_xray_system} for the detailed treatment room layout). To obtain the shift error(SE), we first created a searching pool $S=sgCT+\delta, \delta \in [-1, 1]$, consisting of sgCT and its variances (shifting sgCT within $\pm$ 1mm with a step of 0.1mm), total 21 candidates. Next, the MAE between each candidate sgCT in $S$ and ssCT was calculated. Lastly, by linear search, the sgCT with $\delta_m$ that gave the minimum MAE was identified and the absolute value of $\delta_m$ was defined as SE. We reported the results in Table \ref{shiftrobust} and the model yielded a mean SE of only 0.40 $\pm$ 0.16mm  on average in the sCT robustness test mimicking daily clinic practice, the patient alignment tolerance for H\&N patients is 2-3 mm clinically at our institution.
\renewcommand{\arraystretch}{2}
\begin{table}[h]
  \centering
    \caption{Shift error(SE) (in mm) from all 10 test patients. The lower the SE value is, the more robust the proposed framework is.}
    \label{shiftrobust}    
  \begin{tabular}{ ccccccc }
  \toprule
    {\bf Patient ID} &{\bf 1} &{\bf 2} & {\bf 3} & {\bf 4} &{\bf 5} &{\bf 6}\\
    \hline

    {\bf kV2CTConverter}&{0.37} &{0.56} & {0.66} & {0.25} &{0.44} &{0.34}\\\hline\hline    
 
      {\bf Patient ID} &{\bf 7} &{\bf 8} &{\bf 9} &{\bf 10} &{\bf Average}&{-}\\\hline
      {\bf kV2CTConverter}& {0.63} &{0.26} &{0.31} &{0.22}&{0.40 $\pm$ 0.16} &{-}\\
      \bottomrule
  \end{tabular} 
\end{table} 

\renewcommand{\arraystretch}{2}
\begin{table}
  \centering
    \caption{DVH index difference between the doses calculated on the sCTs and the gCTs using the same plan from the selected patients. The CTV and 4 OARs were considered and the constraints associated with each structure were shown in the (). For each patient, two results were reported as follow: the first row showed the results from kV2CTConverter while the second row was from the primary model only. The full results of the 10 patients were reported in Table \ref{dvhdifffull} in Appendix \ref{secA1}}
  \begin{tabular}{ccccccc}
    \toprule
    \multirow{2}{*}{\bf{Patient ID}}   
      & CTV & CTV & Brain stem& Parotid total & Oral cavity & Mandible\\
      & (D95\%) & (D2\% ) & (D0.01cc)& (Mean) & (Mean) & (D0.03cc)\\
      \midrule\midrule
      \rowcolor{Gray}
     \multirow{2}{*}{\bf 2} & 0.066 &-0.08 & 44cGy & -7.3cGy & 0.1cGy & 2.4cGy\\
     & -4.708 & -0.25 & 148.4cGy & 17cGy & -14.6cGy & 35.1cGy \\\rowcolor{Gray}
     \multirow{2}{*}{ \bf 4} & 0.617 &0.288 & 97.3cGy & 10.1cGy  & 13.4cGy & 13.2cGy \\
       & -6.248 &3.09& -961.6cGy & -8.2cGy &-19.2cGy & 120.4cGy \\\rowcolor{Gray}     
     \multirow{2}{*}{\bf 8} & 0.447 & 2.22 & 53.2cGy & -21.8cGy & 2.8cGy & 329.8cGy \\
     & -0.652 &2.91 & 91.6cGy & 135.6cGy & 57.3cGy & 338.6cGy \\\rowcolor{Gray}
       \multirow{2}{*}{\bf 10} & 1.673 &0.045 & 132.9cGy & 18.7cGy  & 12.9cGy & -6.2cGy \\
    & 7.221 & 4.421 & 212.2cGy & 36.2cGy & 32.7cGy & 15.7cGy \\
    \bottomrule   
  \end{tabular}
  \label{dvhdiff}
\end{table}

\subsection{Ablation Study} 
To investigate if training the secondary model indeed improved the quality of the sCTs, we conducted ablative experiments and reported the results in Table \ref{MAE results} and Table \ref{ctgamma}. From the last row of Table \ref{MAE results}, we observed that adding the secondary model reduced the MAE value by $6.27 \pm 3.71$ HU on average. Although not numerically significant, the visual details in the sCTs trained with both the primary and secondary models, especially in the nasal cavity, brain stem, cochleas, and the surrounding bones were much more distinct than those in the sCTs trained only with the primary model (Fig. \ref{fig:results}\textbf{\textit{d}}). Moreover, the MAE decreased by 38.8\% when only calculated within the head region with the introduction of the secondary model. The observation from the 3D Gamma passing rates of the CT numbers of sCTs reported in Table \ref{ctgamma} also followed the same trend. In addition, The 3D Gamma passing rates of the doses calculated using sCTs and the corresponding DVH index comparison further validated such an observation (Table \ref{dosegamma} and Table \ref{dvhdiff}). This indicated that as a challenging task of sCT generation in this study, which is indispensable to sub-tasks, i.e., locating the target (finding the position of the H\&N region in the CT images) and reconstructing the fine details of the patient anatomies, a practical solution is to train a framework consisting of dual models, each handling one sub-task.   

\section{Discussion}\label{sec3}
\bmhead{kV2CTConverter can effectively reconstruct accurate high-resolution 3D CT from two 2D kV images with arbitrary artefacts}
This is the most prominent novelty in this proposed work. Specifically, to our best knowledge, the proposed framework is the first one to solely take the kV images and their corresponding CToR as the training and testing datasets, without referring to supplementary images such as DRRs. Moreover, different from the proof-of-concept results (i.e. cropped small-size sCT images) reported in the existing approaches\cite{shen2019patient,ying2019x2ct,bayat2020inferring,lei2020deep,maken20232d,gao20233dsrnet,zhang2023xtransct}, we can synthesize full-size sCTs, which can be directly applied in daily clinical practice. In terms of image quality, our framework significantly outperformed previous 2D-to-3D networks\cite{lei2020deep,shen2019patient} by 53.56\% and 62.01\% on average, respectively. We believe that such performance superiority has demonstrated the potential of the proposed framework in clinical applications. 

\bmhead{kV2CTConverter can effectively handle patient setup uncertainty and generate sCT that reflects the real-time patient position shift}
To mimic patient position shift during RT treatment, which commonly happens, we conducted a random shift simulation. The proposed framework was able to generate sCTs reflecting their real-time positions and achieve a minimum SE of $<$0.4mm on average compared with the shifted real patient postions, which is much lesser than the clinic criteria, 2mm or 3mm for H\&N cancer patients. The robustness gain is closely related to the proposed GRSS data augmentation strategy, which not only achieved plenty of data samples for model training but also enabled the framework to cope with random shifts of the patient position accurately. 

\bmhead{The secondary model is essential for fine details recovery in the 3D CT reconstruction.}
From the experiment results reported in Table \ref{MAE results}, Table \ref{ctgamma}, and Table \ref{dosegamma}, we observed that the secondary model indeed boosted the framework performance. Although ViT has outperformed other deep learning-based models in natural image classification and reconstruction, it is still very challenging when employed for tasks related to medical images, as there is a great gap between medical images and natural images. Thus, we proposed a novel dual-model framework to enable the state-of-the-art ViT to be adapted to medical images, in which the primary model was dedicated to identifying the positions of structures of interest, and the secondary model focused on reasoning the 2D-3D relations and reconstructing the voxel-level fine details in 3D CT. Besides, such a dual-model configuration is intuitive, easy to implement, resource-efficient, and can be generalized to other medical imaging modalities, such as MRI, PET, etc. 

\section{Outlook}
From the green boxes shown in Fig. \ref{fig:results}\textbf{\textit{b}}, we noticed that the areas corresponding to the sinuses region, where locates the microscopic soft tissue surrounded by an air cavity with an irregular shape, was not well reconstructed (left green box). Moreover, the brain stem which has an irregular shape and has a similar CT number to the surrounding cerebral hemisphere matters, was not reconstructed with distinct boundaries either (right green box). Intuitively, we can build another model to focus on these two small regions respectively. However, different from the head region considered in the secondary model, where the shape is regular and the position is independent of the helmet, the properties of the two small regions are quite opposite. Thus, it may complicate the framework and double the computing resources. Fortunately, from a purely clinical point of view, those two small regions don't affect a lot: 1) if the tumor is located in those regions, it will either shrink the air cavity or exhibit a clear difference in CT numbers compared to the surrounding tissues, making it easier for the model to distinguish the tumor from nearby organs in either way; 2) if the tumor doesn't appear in those regions, the dosimetric constraints to those regions only possess a low priority when designing the treatment plan. In the future, we will investigate how to further improve the quality of the reconstructed CT regarding those two small regions.

Moreover, if CT with intravenous (IV) contrast are available, it may further improve the identification of the tumor and potentially improve the accuracy of the sCT. However, typically, on the treatment day in clinical practice, the patient will only have the kV images and/or regular CToR whereas the IV contrast is not used. Because the IV contrast will change the CT numbers of the tumor and consequently impact the accuracy of the dose delivery\cite{bae2010intravenous}. Therefore, we don't have kV images and the corresponding CT with IV contrast pairs available for the training of the model. We will leave it for further exploration.

\section{Methods}\label{sec4}
\begin{figure}[!h]
  \centering  
  \includegraphics[width=0.95\linewidth]{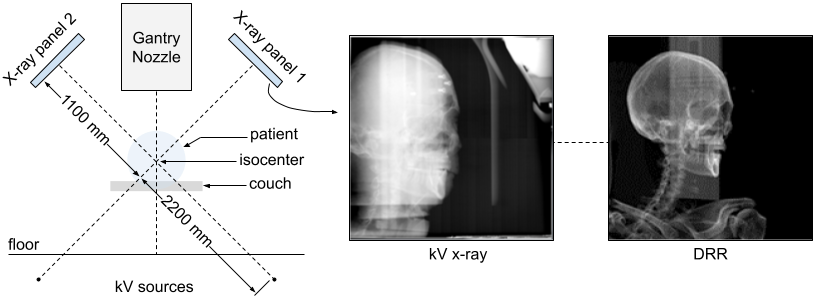}
   \caption{The orthogonal kV x-ray system used for patient alignment at our proton center(left). An exemplary kV image(middle) captured by this system and its corresponding DRR image(right). Comparing to the DRR image, daily-used kV image is often noisy and contains unwanted artifacts from essential medical devices/accessories, such as dental implants or treatment couch attachments.}
   \label{fig:orthogonal_xray_system}
\end{figure}

10 H\&N patients previously treated with IMPT were retrospectively selected for this study. For proton treatment, it is imperative to set up patients such that their anatomy matches the planning CT as closely as possible. This requires 1) precise patient positioning, and 2) ensuring that the patient's anatomy has not diverged from the planning CT. To monitor anatomical changes, verification CT scans are performed regularly (for this study, weekly verification CT scans were performed using the CToR), sometimes leading to a new treatment plan, a process known as ART. Precise patient positioning at our proton center is achieved using an orthogonal 2D kV imaging system. See Figure \ref{fig:orthogonal_xray_system}. This system compares the real-time orthogonal kV images with DRRs generated from the planning CT. After patients are initially aligned, orthogonal kV images are captured and rigidly registered to their corresponding DRRs to determine whether the patient position should be shifted/rotated. After a few iterations of this process, the patient is ready for treatment. As a result of these processes, a large amount of imaging and registration data are generated for each patient over the course of treatment. Of relevance for this study, these data include the initial planning CT, multiple verification CTs (weekly), rigid registration matrices generated by registering the verification CTs to the planning CT (stored as DICOM files), orthogonal kV images (for each treatment), and rigid registration matrices generated by registering the orthogonal kV images to the DRRs. In this study, we took advantage of the fact that patients were typically treated shortly after undergoing a CToR scan. This meant that for each CToR dataset (each patient in this study received a minimum of 3 CToR scans), there was a corresponding pair of orthogonal kV images. In this work, we have investigated the performance of a patient-specific AI model designed to generate sCTs from a pair of 2D orthogonal kV images.

\subsection{Data pre-processing}
The patients in this study were treated with a so-called bolus helmet in place during treatment. The bolus helmet is quite large in terms of volume and was therefore a significant feature in all of the patient imaging. For this reason, two models were developed, the "primary" and "secondary" models. After registered to the same coordinate, the kV images and the CToR image were used for the training of both models as the input and as the ground truth respectively. While the primary model utilized the whole images of the CToR as reference, the secondary model was only provided with the images of the CToR within the head region (see Fig. \ref{fig:model structure}). Over the course of this study, it was discovered that the overall model performance was improved by introducing a secondary model that could focus specifically on the head region. The CToR datasets for each patient had a resolution of $512 \times 512 \times N$, where $N$ was the number of the CT slices along the superior-inferior (S-I) direction (varies from patient to patient). CToR datasets were cropped to two different sizes, one size for the primary model and a smaller size for the secondary model. For the primary model, the CToR images were cropped to size $448 \times 336 \times 384$ to exclude the excessive regions outside the patient BODY with low density (i.e. air) as well as for the purpose of homogenizing the size of the datasets. For the secondary model, the CT images were cropped to size $M \times 224 \times 224$, where $M$ indicates the minimum number of the voxels that covered the head region along the R-L direction, which varies from patient to patient. Finally, the corresponding kV images, initially having resolution of $1024 \times 1024$, were cropped to $1008 \times 1008$ accordingly.

\subsection{Data augmentation}
Converting kV images into a 3D CT is an extremely ill-constrained problem, going from $10^6$ pixels to $10^8$ voxels. For this reason, a novel data augmentation strategy, the so-called \textbf{geometric property-reserved shifting and sampling (GRSS) data augmentation strategy} was proposed. Given the layout of the kV imaging system in the treatment room, as shown in Figure \ref{fig:orthogonal_xray_system}, we noticed that for a shift of the CToR along the S-I direction, the kV images would shift by a factor of $1.5$ in the same direction based on the kV imaging system geometry. Hence, we further augmented the kV-CT pairs by simultaneously moving CT along the S-I direction in steps of 0.1mm (0.15mm for kV images), $\pm$ 5mm in total. In addition to the "property-reserved shifting" step, the shifted datasets were also downsampled. For the primary model, the CT images were downsampled with a factor of 4 along the R-L direction and a factor of 3 in the A-P and S-I directions, respectively. Correspondingly, the two orthogonal kV images were downsampled with a factor of 6 in both dimensions. Likewise, for the secondary model, the CT images were downsampled with a factor of 2 along both the A-P and S-I directions, and the corresponding kV images were downsampled with a factor of $2$ in both dimensions. Thus, a pair of initial CT and its corresponding kV images yields 36 additional shifted and downsampled CT-kV image pairs for the primary model dataset and 4 for the secondary model dataset. This novel method helps to avoid overfitting issues due to the limited number of the training samples. In addition, it allowed for efficient model training since the size of each sample was less than 200 voxels along any direction. Finally, a high-resolution CT of full size ($ 512 \times 512 \times N$), desirable for clinical applications, was obtained by spatially stacking the small-size reconstructed CT generated by both the primary model and secondary model.

\subsection{kV2CTconventer Framework}
The proposed framework, entitled kV2CTconventer (Fig. \ref{fig:model structure}\textbf{\textit{a}}), has dual models. Each has an asymmetric autoencoder-like architecture consisting of an encoder $E_k$ and a decoder $D_r$ with hierarchical ViT blocks as the basic building blocks. The overall architecture of both primary model and secondary model is shown in Fig. \ref{fig:model structure}\textbf{\textit{b}}. Specifically, both models consist of a patch embedding layer (a convolutional layer), an encoder $E_k$, a decoder $D_r$ and a final fully connected layer. The patch embedding layer is used for projecting non-overlapping raw kV image patches to initial high-dimensional feature representations that serves as the input for the encoder $E_k$. Both $E_k$ (Fig. \ref{fig:model structure}\textbf{\textit{c}}) and $D_r$ (Fig. \ref{fig:model structure}\textbf{\textit{d}}) consist of multiple hierarchical ViT blocks, having a pattern of "layer normalization, window-based multi-head attention(W-MHA), layer normalization, multilayer perceptron (MLP), and patch merging/unmerging layer". The W-MHA ((Fig. \ref{fig:model structure}\textbf{\textit{e}})) calculates the attention within the windows only instead of the entire image, thus greatly reducing the computational complexity\cite{liu2021swin,he2022masked}. The patch merging layer in the $E_k$ concatenating nearby $2 \times 2$ patches with a linear merging layer to obtain a hierarchical representation. Likewise, the unmerging layer in the $D_r$ enlarged each patch by a factor of 2 along each dimension through a fully connected layer. Lastly, the final fully connected layer converts from the learned representations to the final output (i.e., the 3D sCT).   

\bmhead{Training protocol} The proposed framework was implemented with the PyTorch deep learning library. Moreover, distributed data parallel (DDP)\cite{li2020pytorch} was employed to minimize memory usage and significantly accelerate the training speed. We used the AdamW optimizer with $\beta_{1} = 0.9$ and $\beta_{2} = 0.999$, and a cosine annealing learning rate scheduler with an initial learning rate of $e^{-7}$ and 20 warm-up epochs. We used smooth L1\cite{girshick2015fast} as the loss function to obtain a smoother loss curve, which is a combination of L1 and L2 loss. The batch size was set to 300 for both the primary model and secondary model. 

\newpage
\bibliography{sn-bibliography}
\newpage
\begin{appendices}

\section{Full Experimental Results}\label{secA1}

\begin{table}
  \centering
    \caption{
    3D Gamma Passing rates (in \%) comparing the CT numbers (HU) of sCTs to gCTs. Three criteria were considered with sCT and gCT as the reference, respectively. In each cell two values were reported, the $1^{st}$ row represented the kV2CTConverter while the $2^{nd}$ represented the primary model only. The last four rows showed the average value over the 10 patients.}
  \begin{tabular}{ccccccc}
    \toprule
    \multirow{2}{*}{\bf{ Patient ID}} 
      & \multicolumn{2}{c}{ 3\%3mm}& \multicolumn{2}{c}{ 3\%2mm}& \multicolumn{2}{c}{ 2\%2mm}\\
      & sCT & gCT& sCT & gCT& sCT & gCT\\
      \midrule
      \rowcolor{Gray}
    \multirow{2}{*}{\bf 1} & 96.78 & 98.32 & 96.54 & 98.21 & 96.54 & 98.20 \\

    & 98.87 & 97.25 & 98.64 & 97.08 & 98.64 & 97.24 \\
     \rowcolor{Gray}\multirow{2}{*}{\bf 2} & 96.91 & 98.94 & 96.46 & 98.81 & 96.46 & 98.80 \\
     & 97.79 & 97.67 & 97.25 & 97.56 & 97.25 & 97.56 \\
     \rowcolor{Gray}
    \multirow{2}{*}{\bf 3} & 93.93 & 99.54 & 93.64 & 99.50 & 93.64 & 99.50 \\
    & 94.19 & 97.87 & 93.89 & 97.79 & 93.89 & 97.79 \\
        \rowcolor{Gray}\multirow{2}{*}{\bf 4} & 98.15 & 98.85 & 97.80 & 98.69 & 97.80 & 98.70 \\
        & 98.33 & 97.41 & 97.80 & 97.28 & 97.80 & 97.28 \\
     \rowcolor{Gray}\multirow{2}{*}{\bf 5} & 97.88 & 99.09 & 97.34 & 98.94 & 97.34 & 98.94 \\
     & 97.75 & 98.26 & 97.25 & 98.02 & 97.25 & 98.02 \\
    \rowcolor{Gray}\multirow{2}{*}{\bf 6} & 98.47 & 95.38 & 98.08 & 94.97 & 98.07 & 94.97 \\
    & 97.31 & 95.14 & 96.80 & 95.21 & 96.80 & 94.99 \\
        \rowcolor{Gray}\multirow{2}{*}{\bf 7} & 98.71 & 99.75 & 98.61 & 99.72 & 98.61 & 99.72 \\
        & 98.12 & 99.61 & 98.02 & 99.54 & 98.02 & 99.54 \\
     \rowcolor{Gray}\multirow{2}{*}{\bf 8} & 99.32 & 98.95 & 99.14 & 98.80 & 99.14 & 98.80 \\
     & 99.21 & 98.19 & 99.03 & 98.08 & 99.03 & 98.08 \\
    \rowcolor{Gray}\multirow{2}{*}{\bf 9} & 93.81 & 98.35 & 92.97 & 98.15 & 92.97 & 98.15 \\
    & 94.56 & 92.44 & 93.57 & 92.21 & 93.57 & 92.21 \\
    \rowcolor{Gray}\multirow{2}{*}{\bf 10} & 99.97 & 99.81 & 99.87 & 99.52 & 99.85 & 99.50 \\
    & 96.76 & 97.81 & 95.13 & 97.67 & 94.62 & 97.55 \\\hline
    \rowcolor{Gray}   & 97.39 $\pm$ 2.09  & 98.70 $\pm$ 1.27 &   97.05 $\pm$ 2.24   &98.53 $\pm$ 1.36 & 97.04 $\pm$ 2.24 &   98.53 $\pm$ 1.36 \\
    \rowcolor{Gray} \multirow{-2}{*}{\bf{kV2CT}}& \multicolumn{2}{c}{ 98.05 $\pm$ 1.82}& \multicolumn{2}{c}{ 97.79 $\pm$ 1.96}& \multicolumn{2}{c}{ 97.79 $\pm$ 1.96} \\
    \multirow{2}{*}{\bf{Primary}}   & 97.29 $\pm$ 1.69&   97.17 $\pm$ 2.00&   96.74 $\pm$ 1.91&   97.04 $\pm$ 2.01 &   96.69 $\pm$ 1.97 &   97.03 $\pm$ 2.02 \\
    & \multicolumn{2}{c}{ 97.23 $\pm$ 1.80}& \multicolumn{2}{c}{ 96.89 $\pm$ 1.91}& \multicolumn{2}{c}{ 96.86 $\pm$ 1.95} \\
    \bottomrule    
    \bottomrule
  \end{tabular}
  \label{ctgammafull}
\end{table}

\begin{table}
  \centering
    \caption{3D Gamma Passing rates (in \%) of the doses calculated on the sCTs and the gCTs using the same plan. Three criteria were considered with the dose calculated on the sCT and gCT as the reference, respectively. In each cell two values were reported, the $1^{st}$ row represented the kV2CTConverter while the $2^{nd}$ represented the primary model only. The last four rows showed the average value over the 10 patients.}
  \begin{tabular}{ccccccc}
    \toprule
    \multirow{2}{*}{\bf{Patient ID}} &
       \multicolumn{2}{c}{ 3\%3mm}& \multicolumn{2}{c}{ 3\%2mm}& \multicolumn{2}{c}{ 2\%2mm}\\
      & sCT & gCT& sCT & gCT& sCT & gCT\\
      \midrule
      \rowcolor{Gray}
    \multirow{2}{*}{\bf 1} & 99.99 & 99.96 & 99.94 & 99.85 & 99.66 & 99.51 \\     & 93.18 & 90.73 & 88.36 & 86.62 & 85.45 & 83.44 \\
     \rowcolor{Gray}\multirow{2}{*}{\bf 2} & 99.75 & 99.83 & 99.47 & 99.49 & 98.91 & 98.96 \\& 97.98 & 97.05 & 95.56 & 94.72 & 93.14 & 92.08 \\
    \rowcolor{Gray}\multirow{2}{*}{\bf 3} & 99.36 & 99.67 & 98.54 & 98.95 & 97.28 & 97.70 \\& 90.39 & 89.77 & 86.30 & 86.14 & 83.14 & 82.93 \\
        \rowcolor{Gray}\multirow{2}{*}{\bf 4} & 99.93 & 99.89 & 99.66 & 99.55 & 99.02 & 98.89 \\& 96.04 & 93.25 & 92.41 & 90.21 & 89.87 & 87.38 \\
     \rowcolor{Gray}\multirow{2}{*}{\bf 5} & 99.99 & 99.99 & 99.94 & 99.97 & 99.82 & 99.83 \\& 97.89 & 98.21 & 97.59 & 97.97 & 97.59 & 97.97  \\
    \rowcolor{Gray}\multirow{2}{*}{\bf 6} & 99.22 & 98.63 & 96.99 & 96.08 & 95.46 & 94.33 \\& 93.05 & 95.19 & 88.18 & 90.96 & 84.92 & 87.75 \\\rowcolor{Gray}
        \multirow{2}{*}{\bf 7} & 99.94 & 99.93 & 99.55 & 99.56 & 98.04 & 98.14 \\& 93.17 & 87.84 & 87.75 & 82.78 & 84.15 & 79.05 \\\rowcolor{Gray}
     \multirow{2}{*}{\bf 8} & 99.23 & 99.55 & 98.40 & 98.81 & 97.29 & 97.70 \\& 95.05 & 94.27 & 91.74 & 91.30 & 88.66 & 88.17 \\\rowcolor{Gray}
    \multirow{2}{*}{\bf 9} & 99.31 & 99.09 & 97.34 & 97.22 & 94.70 & 94.65 \\& 92.94 & 93.39 & 87.13 & 87.63 & 82.26 & 82.70 \\\rowcolor{Gray}
    \multirow{2}{*}{\bf 10} & 99.97 & 99.95 & 99.66 & 99.57 & 98.56 & 98.49 \\& 91.39 & 90.67 & 88.21 & 88.54 & 86.74 & 85.06 \\\hline
    \rowcolor{Gray}   & 99.67 $\pm$ 0.34 &   99.65  $\pm$ 0.45 &   98.95 $\pm$ 1.08 &   98.91 $\pm$ 1.27&   97.87 $\pm$ 1.71&    97.82 $\pm$ 1.89 \\    
    \rowcolor{Gray}\multirow{-2}{*}{\bf{kV2CT}}& \multicolumn{2}{c}{ 99.66 $\pm$ 0.39}& \multicolumn{2}{c}{ 98.93 $\pm$ 1.15}& \multicolumn{2}{c}{ 97.85 $\pm$ 1.76} \\
    \multirow{2}{*}{\bf{Primary}}   & 94.11 $\pm$ 2.57 &   93.04 $\pm$ 3.30 &   90.32 $\pm$ 3.84 &   89.69 $\pm$ 4.39 &   87.59 $\pm$ 4.83 &   86.65 $\pm$ 5.39 \\
    & \multicolumn{2}{c}{ 93.57 $\pm$ 2.93}& \multicolumn{2}{c}{ 90.01 $\pm$ 4.03}& \multicolumn{2}{c}{ 87.12 $\pm$ 5.00} \\
    \bottomrule    
    \bottomrule
  \end{tabular}
  \label{dosegammafull}
\end{table}

\begin{table}
  \centering
    \caption{DVH index difference between the doses calculated on the sCTs and the gCTs using the same plan. The CTV and 4 OARs were considered and the constraints associated with each structure were shown in the (). For each patient, two results were reported, the first row showed the results from kV2CTConverter, while the the second row was from the primary model only.}
  \begin{tabular}{ccccccc}
    \toprule
    \multirow{2}{*}{\bf{Patient ID}}   
      & CTV & CTV & Brain stem& Parotid total & Oral cavity & Mandible\\
      & (D95\%) & (D2\% ) & (D0.01cc)& (Mean) & (Mean) & (D0.03cc)\\
      \midrule\midrule
      \rowcolor{Gray}
    \multirow{2}{*}{\bf 1} & 1.394 & -0.97& 27.1cGy & 1.7cGy & 43.1cGy & - \\ 
    & -66.381 &-2.96 & 2525cGy & 19.9cGy & -140.7cGy & - \\ 
   \rowcolor{Gray}
     \multirow{2}{*}{\bf 2} & 0.066 &-0.08 & 44cGy & -7.3cGy & 0.1cGy & 2.4cGy\\
     & -4.708 & -0.25 & 148.4cGy & 17cGy & -14.6cGy & 35.1cGy \\\rowcolor{Gray}
    \multirow{2}{*}{\bf 3} & -3.25 &0.172& -11cGy & - & - & - \\
    & -32.466 &2.5 & 53cGy & - & - & - \\\rowcolor{Gray}
       \multirow{2}{*}{ \bf 4} & 0.617 &0.288 & 97.3cGy & 10.1cGy  & 13.4cGy & 13.2cGy \\
       & -6.248 &3.09& -961.6cGy & -8.2cGy &-19.2cGy & 120.4cGy \\\rowcolor{Gray}
     \multirow{2}{*}{\bf 5} & 0.229 &-0.045 & 69.3cGy & 0.6cGy & 0 & 32cGy \\
     & 0.45 &-0.21 & 112.2cGy & 3.8cGy & 1.2cGy & 38cGy \\\rowcolor{Gray}
    \multirow{2}{*}{\bf 6} & 0.35 &-0.69 & 210.8cGy & - & - & - \\
    & -5.832 &1.56 & -1078.9cGy & -  & - & - \\\rowcolor{Gray}
        \multirow{2}{*}{\bf 7} & -1.786 &0.31 & 0 & 0 & -49 & -3cGy \\
        & -15.927 &0.33 & 0 & 0 & -62.7cGy & 128.8cGy \\\rowcolor{Gray}
     \multirow{2}{*}{\bf 8} & 0.447 & 2.22 & 53.2cGy & -21.8cGy & 2.8cGy & 329.8cGy \\
     & -0.652 &2.91 & 91.6cGy & 135.6cGy & 57.3cGy & 338.6cGy \\\rowcolor{Gray}
    \multirow{2}{*}{\bf 9} & 0.094 &-2.76 & -7.6cGy & -105.2cGy & 15.7cGy & 22.8cGy \\
    & 2.583 &4.44& 188.1cGy & 204cGy & 105.3cGy & -21.4cGy \\\rowcolor{Gray}
    \multirow{2}{*}{\bf 10} & 1.673 &0.045 & 132.9cGy & 18.7cGy  & 12.9cGy & -6.2cGy \\
    & 7.221 & 4.421 & 212.2cGy & 36.2cGy & 32.7cGy & 15.7cGy \\\rowcolor{Gray}
    \bottomrule   
  \end{tabular}
  \label{dvhdifffull}
\end{table}
\end{appendices}
 \end{document}